\documentclass[letterpaper, 12pt]{article} 
\usepackage[top=2cm, bottom=2cm, left=2cm, right=2cm]{geometry}
\usepackage{setspace}
\usepackage[backend=bibtex,sorting=none, style=numeric-comp,maxbibnames=99]{biblatex}
\usepackage{amsmath,amssymb}
\usepackage[colorlinks=true,bookmarks=false,citecolor=blue,urlcolor=blue]{hyperref} 
\usepackage{authblk}
\usepackage{abstract}
\usepackage{graphicx}
\usepackage{bm}

\title{\textbf{Shape optimization for high efficiency metasurfaces: theory and implementation}}

\author[1,*]{P. Dainese}
\author[1]{L. Marra}
\author[2]{D. Cassara}
\author[2]{A. Portes}
\author[2]{J. Oh}
\author[1]{J. Yang}
\author[2]{A. Palmieri}
\author[1]{J.R. Rodrigues}
\author[2]{A.H. Dorrah}
\author[2]{F. Capasso}

\affil[1]{Corning Research and Development Corporation, 184 Science Center Dr, Painted Post, NY 14870, USA}
\affil[2]{Harvard John A. Paulson School of Engineering and Applied Sciences 1, Cambridge, MA 02138, USA}

\affil[*]{\href{mailto:dainesep@corning.com}{dainesep@corning.com}}
\date{}
\begin{document}

\maketitle

\begin{abstract}
	
Complex non-local behavior makes designing high efficiency and multifunctional metasurfaces a significant challenge. While using libraries of meta-atoms provide a simple and fast implementation methodology, pillar to pillar interaction often imposes performance limitations. On the other extreme, inverse design based on topology optimization leverages non-local coupling to achieve high efficiency, but leads to complex and difficult to fabricate structures. In this paper, we demonstrate numerically and experimentally a shape optimization method that enables high efficiency metasurfaces while providing direct control of the structure complexity. The proposed method provides a path towards manufacturability of inverse-designed high efficiency metasurfaces.

\end{abstract}

\newpage

\section{Introduction}

Wavefront shaping using metasurfaces has attracted significant scientific and technological interest in recent years \cite{yu2014flat, kildishev2013planar, arbabi2015dielectric, kuznetsov2016optically, lalanne2017metalenses, kuznetsov2024roadmap}. The ability to control many degrees of freedom of an incoming beam, including phase, amplitude, polarization, and dispersion has led to a number of demonstrations in areas ranging from imaging, polarization optics, communications, atom trapping, optical computing and image processing, nonlinear optics and others \cite{chen2018broadband, arbabi2018full, rubin2019matrix, dorrah2022tunable, oh2022adjoint, oh2023metasurfaces, chen2021edge, cordaro2023solving, jammi2024three, tseng2022vacuum}. This versatility stems from a large design space enabled by engineering meta-atoms with different geometrical shapes and materials readily available in nano-fabrication, which allows tapping into different mechanisms such as Mie scattering, waveguide propagation phase, Pancharatnam-Berry phase, guided mode resonances, and more broadly Bloch-mode engineering \cite{lalanne2006optical, sell2017large}. Despite successful demonstrations, designing metasurfaces is still a significant challenge due to the complex nature of such physical mechanisms, particularly when sub-wavelength and often non-periodic arrangement leads to strong non-local coupling between meta-atoms \cite{lalanne2017metalenses,campbell2019review,kuznetsov2024roadmap}. 

\begin{figure} [t!]
	\centering
	\includegraphics{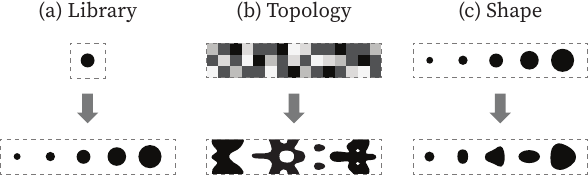}
	\caption{ Metasurface design methodologies. In the library method (a), the phase and amplitude response of pre-determined meta-atoms is calculated as a function of its parameters (e.g. diameter for a circular pillar). The response is then used to create devices by dividing the surface in unit cells, each containing a specific element of the library. In topology optimization (b), the design domain is typically initialized with a randomized refractive index distribution, and the adjoint algorithm is used to converge towards a final binarized geometry. In shape optimization (c), only smooth variations in existing geometrical features are allowed.}
	\label{fig:methods}
\end{figure}

The most common design approach is based on libraries of meta-atoms (Figure \ref{fig:methods}a), using either parametrized shapes or free-form meta-atoms \cite{stork1991artificial,chen1996diffractive,lalanne1998blazed,yu2011light,arbabi2015dielectric,chen2018broadband,whiting2020meta}. The simplicity of this method relies on the assumption that each meta-atom is placed on a unit cell with periodic boundary conditions, allowing fast computation of its response using numerical methods. Metagratings, metalenses and more complex devices have been created using libraries. However, except for truly periodic arrays, this assumption is necessarily broken in real devices and pillar-to-pillar interaction imposes a fundamental limitation to performance by perturbing the realized phase and intensity profile, making it difficult to achieve high efficiency in general. Another limitation of the library method is that the unit cell response is computed for a specific angle of incidence, typically normal, and for a specific polarization. In general, however, an incoming beam may contain a spectrum of incidence angles, as is the case even for a simple metalens illuminated with non-collimated light, and in more general holograms used for vector beam generation, spatial mode multiplexing and many others.

To overcome these challenges, topology optimization based on the adjoint method has been proposed ~\cite{jensen2011topology, molesky2018inverse,sell2017large,chung2020high,sang2022toward}. In this approach, an initial random refractive index distribution iteratively converges to a final binary, i.e. manufacturable, structure (Figure \ref{fig:methods}b). Rigorous electromagnetic simulation of the full device ensures that the interaction between meta-structures is fully considered. Furthermore, the adjoint formulation enables computing gradients of a given figure of merit with respect to the refractive index at every pixel in the domain using only two electromagnetic simulations (forward and adjoint simulations), critical for optimization problems with such high dimensionality. Despite successful designs with very high efficiencies, controlling the complexity of the final structure is a significant challenge, particularly for large scale manufacturing. During the optimization process, the device structure evolves according to the topological derivative at each iteration, and often leads to features that are difficult to control in patterning and etching processes, such as the appearance of sharp boundaries and small features such as islands of materials, small holes and small gaps between structures. These challenges have stimulated recent research to incorporate manufacturing robustness in the design process \cite{jensen2011topology, wang2019robust,sang2022toward, kuznetsov2024roadmap}, either external to the adjoint formulation such as applying structural blurring every so often during optimization, or including penalty terms in the figure of merit to \textit{guide} the topological derivative in a way to minimize these issues.

The adjoint formulation can also be used to extract gradients of the figure of merit with respect to boundary shifts at the interface between two materials. In this case, the topology of the structure is \textit{unchanged} during the optimization, and only deformations of the existing boundaries occur (Figure \ref{fig:methods}c). This approach has been applied to optimize the shape of planar photonic devices, such as waveguide splitters, crossing and bends in photonic crystal waveguides \cite{liu2013compact, lalau2013adjoint,lebbe2019robust}. Leveraging boundary gradients for metasurface design has had limited investigation, with only a few examples where the basic meta-atom shape remained unchanged but their sizes were optimized. This was applied to design metalenses by optmizing the side lengths of rectangular pillars \cite{mansouree2021large}, the radii of circular pillars \cite{mansouree2021large}, as well as to design metagratings where the semi-axis of elliptical pillars were optimized \cite{gershnabel2022reparameterization,zhou2024large, zhou2024inverse}. In this paper, we generalize this approach and investigate a \textit{shape} optimization method that achieves efficiencies higher than the library method by fully considering pillar-to-pillar interaction, while providing greater control over the structure complexity compared to topology optimization. The initial shape of each meta-atom in the device is \textit{smoothly} deformed throughout the process, with the shape complexity controlled through a Fourier decomposition of the adjoint boundary gradients. Direct control of all boundaries naturally incorporates fabrication constraints such as minimum feature size and minimum gap, and by construction, excludes the appearance of holes. Similarly to topology optimization, shape optimization can be applied to any kind of metasurface devices, can handle any input and target field distributions, as well as include multiple objectives. The paper is organized as follows: we outline the formulation in Section \ref{section:formulation}, and then apply the shape optimization method to design several high efficiency metagratings and metalenses in Section \ref{section:numerical}. Experimental results are presented in Section \ref{section:experiment}, and we draw conclusions in Section \ref{section:conclusions}.

\section{Shape optimization formulation}
\label{section:formulation}

\begin{figure}[h!]
	\centering
	\includegraphics{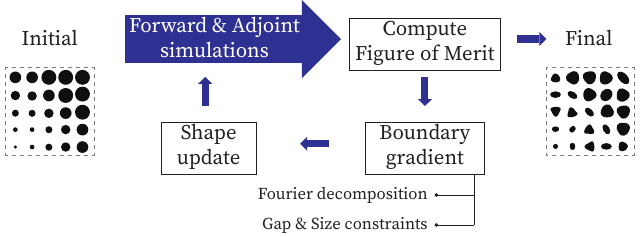}
	\caption{Flow chart of the optimization process. The design domain is initialized with a given set of meta-atoms, for example with a library-based metasurface. Forward and adjoint simulations are performed and the figure of merit computed. If the figure of merit satisfies the target or if it has converged the optimization stops. If not, the shifting boundary gradients are calculated from the simulated fields, followed by Fourier decomposition and possibly additional fabrication constraints. The shape is then deformed according to the gradients and a new simulation is performed. The process repeats until convergence.}
	\label{fig:flowchart}
\end{figure}

Topology and shape optimization in photonics are both inverse design techniques based on the adjoint method~\cite{jensen2011topology,molesky2018inverse}. A flowchart of  the shape optimization method is shown in Figure \ref{fig:flowchart}. The design domain is initialized with a given set of meta-atoms,  for example using a uniform array of circular pillars, a library based device, or even a random distribution of pillars. At each iteration, two electromagnetic simulations are performed, a forward and an adjoint simulation, from which shape gradients for \textit{all} pillars are computed for a given figure of merit. If at any iteration the figure of merit of the device reaches a desired target or if it converges to a local maxima, the optimization stops. Otherwise, the gradients are used to update the shapes and continue on to the next iteration. Before updating the shapes, the gradients are first decomposed in Fourier basis and further fabrication constraints can be implemented. 

The mathematical formulation of the adjoint shape gradient and subsequent Fourier decomposition is outlined here, and more details can be found in Appendix \ref{appendix:math}. As illustrated in Figure \ref{fig:fourier}a, in the forward simulation the incident field propagates through the metasurface, while in the adjoint simulation the target field propagates backwards. In this example, the forward field is simply a plane wave incident at normal direction, and the target field is a plane wave propagating at a certain deflection angle. Through a reciprocity argument  \cite{landau2013electrodynamics}, the forward $(E)$ and adjoint $(E_a)$ fields at the surface of each pillar can be used to compute variations in the figure of merit $(\eta)$ due to an arbitrary (but small) boundary shift $u_{\perp}$:

\begin{equation}
	\label{eq:delta_eta_general}
	\delta \eta = \frac{\omega\delta\epsilon}{2P} Re\left[jF^* \int{u_{\perp} (E_{\parallel}\cdot E_{a,\parallel} + \frac{1}{\epsilon_{1}\epsilon_{2}}D_{\perp}\cdot D_{a,\perp}) \ da} \right],
\end{equation}

\noindent where $\omega$ is the optical frequency, $\delta\epsilon = \epsilon_{2} - \epsilon_{1}$ is the difference in dielectric permittivity between the meta-atom and the surrounding medium, and $P$ is a normalization power. The subscripts $\parallel$ and $\perp$ represent the tangential and normal components of the fields ($F$ is related to the device efficiency, $\eta=|F|^2$). The integration in equation \ref{eq:delta_eta_general} is performed on every pillar's surface. Since we wish to retain vertical pillars for top-down fabrication, we enforce the boundary displacement $u_{\perp}$ to be uniform along $z$, and allow only variation along the cross-sectional boundary. With that, we can explicitly write the efficiency change as an integral along the pillar closed cross-sectional boundary as:

\begin{equation}
	\label{eq:delta_eta}
	\delta \eta = \frac{\omega\delta\epsilon}{2P} \oint{u_{\perp} g \ ds},
\end{equation}

\noindent where the gradient function $g$ is defined accordingly as:

\begin{equation}
	\label{eq:gradient}
	g = Re\left[jF^* \int{(E_{\parallel}\cdot E_{a,\parallel} + \frac{1}{\epsilon_{1}\epsilon_{2}}D_{\perp}\cdot D_{a,\perp}) \ dz} \right].
\end{equation}

\begin{figure}
	\centering
	\includegraphics{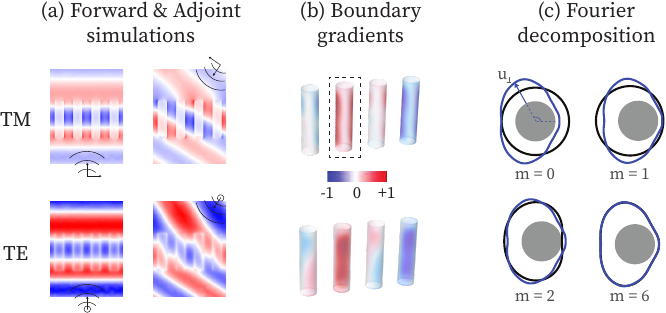}
	\caption{Details of the shape optimization algorithm. The forward and adjoint fields are computed in the design domain, and are then used to calculate a gradient function $g$ on the surface of all existing meta-atoms. (a) and (b) show the fields and the derived boundary gradients for TM and TE polarizations. The gradient function determines the magnitude and direction that each point on the surface must be displaced to increase the device efficiency. In (c), the resulting shape gradient is plotted (blue) along with its Fourier decomposition (black) for the pillar highlighted in (b). One can observe that the gradient containts significant contributions up to order $m = 6$.}
	\label{fig:fourier}
\end{figure}

From the expression of $\delta \eta$ in equation \ref{eq:delta_eta}, it is clear that the efficiency change is always positive if we choose the boundary deformation $u_{\perp}$ to follow the gradient function $g$, i.e., by choosing $u_{\perp} = h sign(\delta \epsilon)  g$, where $h$ is a scaling factor. Mathematically, this is simply the projection of the gradient function on itself, as the integral represents the inner product in the closed boundary domain. The term $(E_{\parallel}\cdot E_{a,\parallel} + \frac{1}{\epsilon_{1}\epsilon_{2}}D_{\perp}\cdot D_{a,\perp})$ in equation \ref{eq:gradient} explicitly determines how the forward and adjoint fields determine the shape gradient $u_{\perp}$. This term is plotted on the pillars' surface in Figure \ref{fig:flowchart}b, with an arbitrary -1 to 1 color scale. There are several aspects worth discussing. First, clearly some pillars tend to increase in size (red colors) while others tend to shrink (blue colors), creating a size gradient that eventually will form the desired metagrating. Second, the gradients are not symmetric along the circular boundary, indicating that the shapes will tend to deviate from simple circles. Physically, this asymmetry arises from the field discontinuities at regions where the linearly polarized incident field is normal to the pillar surface. Finally, as a consequence of the last point, TE and TM polarizations create different gradients, and tend to deform the pillars differently. This is quite clearly seen in the second pillar from the left, where the gradients seem somewhat rotated by $90\deg$. This difference in gradients means that it is more challenging to optimize a structure to simultaneously diffract both TE and TM polarizations with high efficiencies. This argument can be extended to problems with multiple objectives, for example optimizing a multi-wavelength device, where each wavelength tends to create different forward and adjoint field distributions, and therefore different shape gradients.  

The gradients discussed in Figure \ref{fig:fourier}b continuously deform the shape at each iteration. As discussed, it is essential for manufacturability that certain constraints are observed. The deformation function $u_{\perp}=g$ may in general be very complicated, leading to complex shapes after many iterations. Since we are dealing with a closed boundary, any function can be expanded in terms of an appropriate basis defined in such domain. We chose Fourier as it easily allows restriction to smooth round structures. Decomposing the gradient function $g$ in a Fourier series, the boundary displacement is expressed as: 

\begin{equation}
	\label{eq:u_fourier}
	u_{\perp} = h\, \text{sign} (\delta \epsilon) \left( \frac{a_0}{2} + \sum_{m = 1}^{\infty} a_m \text{cos} \, m \theta + b_m \text{sin}\, m \theta \right),
\end{equation}

\noindent where $(a_m,b_m)$ are the expansion coefficients. With that, the efficiency variation in equation \ref{eq:delta_eta} is then written as:

\begin{equation}
	\label{eq:eff_fourier}
	\delta \eta = \frac{\omega\delta\epsilon}{4P}\, h\, s\, \text{sign} (\delta \epsilon) \left[ \frac{a_0^2}{2} + \sum_{m=1}^{\infty}(a_m^2 + b_m^2) \right],
\end{equation}

Note that because we use a set of orthogonal basis (Fourier in this example), each coefficient appears squared in the brackets. This means that each term of the expansion independently contributes to increasing the metasurface efficiency. We can freely choose to drop certain coefficients and still, the remaining ones will always push the efficiency upwards (or remain unchanged if a local optimum has been reached, but never reduce the efficiency). For example, one might choose to keep only the zero-order ($m=0$) coefficient, ensuring that the pillars remain circular throughout the optimization process (of course their diameters are allowed to change). Controlling the Fourier order $m$ gives explicit control of the trade-off between performance and complexity at every iteration. To illustrate this point, we show in Figure \ref{fig:fourier}c the shape gradient and its decomposition for one of the pillars (highlighted in Figure \ref{fig:fourier}b). As one can see, restricting to zero-order $m=0$ simply increases the pillar diameter. Adding the first-order term $m=1$ introduces a shift in the center position of the pillar, while the second-order $m=2$ creates a certain ellipticity. Finally, adding up to the sixth $m=6$ order is sufficient to closely represent the full gradient function for this particular case. In the numerical examples presented in Section \ref{section:numerical}, we show various devices designed with different Fourier orders, illustrating the ability control the device complexity. 

Another important constraint for fabrication is to ensure that the gap between particles is not too small, which can severely intensify the issue of gap-dependent etch rate, leading to under- or over-etching regions and sidewall angle variations. In the shape optimization method, at every iteration we have the explicit boundary coordinates $\bm{r_i}$ before deformation and $\bm{r_{i+1}} = \bm{r_{i}}+u_{\perp} \bm{n_{i}}$ after deformation, where $\bm{n_{i}}$ is the normal unit vector at a given point on the boundary. In the simplest form, we can limit the scaling factor $h$ so that the deformed boundary always respects a target $gap$ to the unit cell boundary. Finally, it is also important to limit the minimum feature size to avoid challenges with patterning, etching and pillars falling over, and again, this becomes straightforward given direct knowledge of the boundary coordinates. 

\section{Numerical results}
\label{section:numerical}

To illustrate the method, we applied shape optimization to design several metagratings and metalenses at $1.55 ~\mu m$ operating wavelength, all based on $1 ~\mu m$ tall amorphous silicon (aSi) pillars on a glass substrate. In every example discussed here, we imposed minimum gap between pillars of 90 nm and minimum feature size of 80 nm. Furthermore, we targeted polarization insensitive operation and therefore simultaneously optimized for TE and TM incident polarizations. In the first example, a meta-grating was designed to deflect light at $51^{\circ}$. The initial structure was created based on the library approach, and contains 4 circular pillars, each on a 500 nm unit cell. In the library method, the pillars' diameters are chosen so that they impart a phase profile with steps of $2\pi/N$ ($N=4$ in this example). The phase imparted by the first unit cell is, however, arbitrary, as different values simply represent different global phases. One can therefore freely choose the diameter of the first pillar as long as subsequent diameters correspond to a $2\pi/N$ phase shift. Despite nominally imparting equivalent phase profiles, these meta-gratings with different pillar sizes exhibit distinct non-local interaction, and may perform very differently. This is exactly the case observed here for the $51^{\circ}$, in which certain choices of the first pillar diameter create a resonance that severely reduces the first order diffraction efficiency for TE light. We applied  shape optimization to two metagratings differing only by the choice of the first pillar diameter, one exhibiting low efficiency due to the appearance of a resonance for TE, and another with the highest efficiency for this library (which was found after sweeping the first pillar diameter for all values available in the library from 100 nm to 410 nm). 

\begin{figure}[h!]
	\centering
	\includegraphics{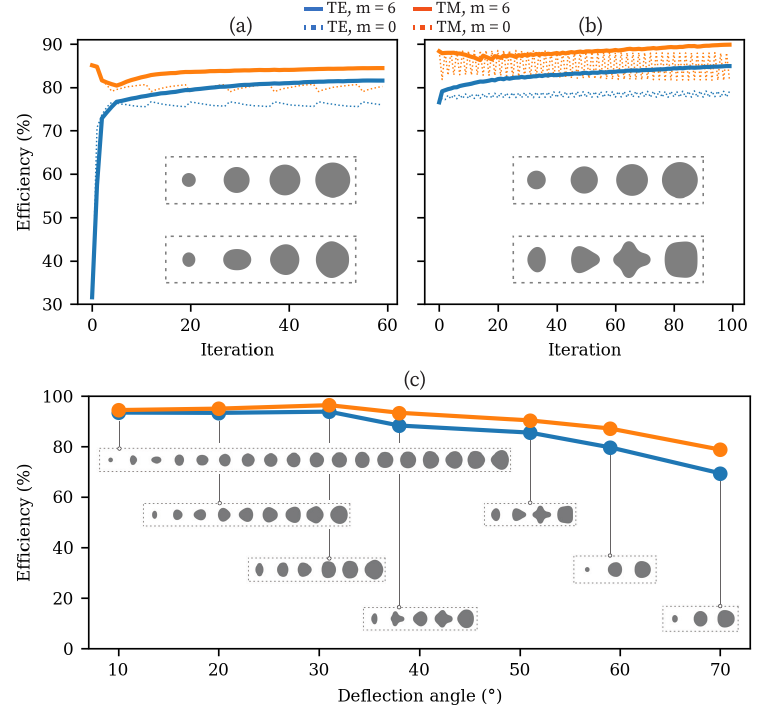}
	\caption{Optimization of a dual-polarization 51-degree beam deflector. Shape optimization was applied to two library designs differing only by the choice of the first pillar diameter: in (a), the initial design exhibits low efficiency due to the appearance of a resonance for TE polarization while in (b) the initial structure exhibited the highest efficiency possible for such library. The curves show the evolution of the absolute efficiency at each iteration, where dashed curves represent restricting the gradient function to $m = 0$ Fourier order (i.e. maintaining circular shape), while solid line allows up to $m = 6$. The final structures for both Fourier orders are shown in the insets. In (c), shape optimization was applied to design meta-gratings from 0 to 70 degrees (final geometries shown as insets). The color legend on top of the figure applies to all plots.}
	\label{fig:gratings_sim}
\end{figure}

The initial library design for the \textit{resonant} metagrating has first order diffraction efficiency of only about 30\% for TE polarization (see iteration = 0 in Figure \ref{fig:gratings_sim}a), while it is as high as 85\% for TM polarization. Such low performance for TE cannot be predicted from the performance of the individual meta-atoms, as all values of pillar diameters in the library are highly transmissive (above 80\%), and of course their respsonse is polarization independent by symmetry. We then applied shape optimization and observed that the performance is significantly improved as shown in  Figure \ref{fig:gratings_sim}a. The enhancement was observed either when the Fourier order is restricted to $m=0$ (dashed lines, final geometry in upper inset) or when we allowed up to $m=6$ (solid lines, final geometry in lower inset). Note that in both cases, at the beginning of the optimization over the first few iterations, the improvement in TE polarization comes at the expense of the TM efficiency. This initial degradation in the TM efficiency could not be recovered with the $m = 0$ case, demonstrating a clear limitation in restricting the shapes to circles. With higher Fourier orders, eventually the TM efficiency recovers leading to a device with efficiencies of 82\% and 84\% for TE and TM, respectively. Despite allowing higher Fourier order, the final shapes are still smooth and respect the gap and minimum feature constraints. In the second case shown in Figure \ref{fig:gratings_sim}b, the initial structure was the highest average TE/TM efficiency obtained with this library. Despite that, the initial efficiency of 76\% for TE was still significantly improved to 85\%, with no degradation on the TM efficiency (which actually slightly improved from 88\% to 90\%). This example illustrates that the shape optimization can be used to eliminate resonances that are not possible to predict from the library alone, as well as to push the performance beyond the highest achievable with such library. Degradation in efficiency due to non-local effects is not specific to this $51^{\circ}$ metagrating, and was observed in other deflection angles from 10 to 70 degrees, in some cases TE and in other cases for TM polarizations. Furthermore, a choice of initial diameter that leads to relatively good performance for one deflection angle is not necessarily the best choice for another angle, illustrating another challenge of relying solely on the library approach. We designed various other metagratings using shape optimzation from angles varying between 10 and 70 degrees, with the final efficiencies and shapes shown in Figure \ref{fig:gratings_sim}c. For example, the 70 degrees metagrating showed average TE/TM efficiency of 74\%.

\begin{figure}
	\centering
	\includegraphics{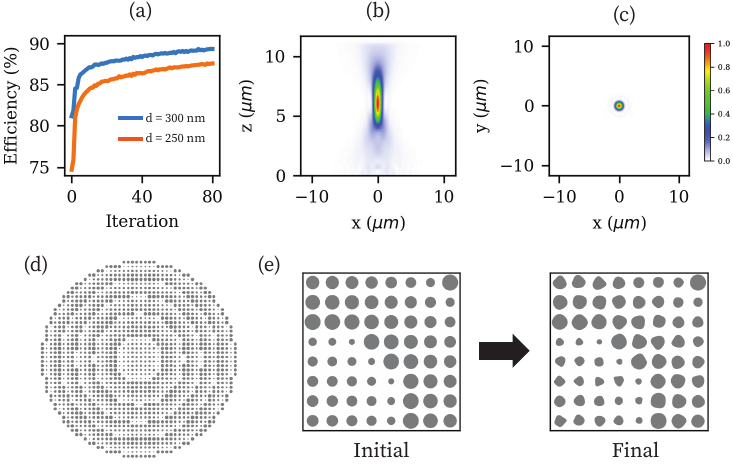}
	\caption{(a) evolution of the absolute efficiency for two different initial structures; cuts of the z-component of Poynting vector in the xz plane in (b) and in the xy plane in (c). The color scale shown in (c) also applies for (b). The full metalens structure is shown in (d), and a zoomed region at the metalens center before and after shape optimizaiton is shown in (e).}
	\label{fig:lens}
\end{figure}

As mentioned, the shape optimization method is not limited to metagratings and can be applied to any general metasurface. To illustrate this, we designed a metalens operating at $1.55 ~\mu m$ and the results are shown in Figure \ref{fig:lens}. The optimization domain was again initialized with a structure based on a library of aSi circular pillars on a glass substrate. The library unit cell is $500 ~nm$ and the pillars range from $100$ to $410 ~nm$ in diameter, with height of $1 ~\mu m$. The lens was designed to impart a phase transformation $\phi = k(\sqrt{r^2 + f^2}-f)$ with focus of $5.3 ~\mu m$ over a total diameter of $2r = 23 ~\mu m$, nominally resulting in a numerical aperture of $NA = sin[tan^{-1}(r/f)] = 0.92$. However, the lens was illuminated with a Gaussian beam with waist radius $w = 5.2 ~\mu m$, which focused at $f = 5.3 ~\mu m$ limits the NA of the system to approximately $NA = 0.83$. We optimized two metalenses that differ only by a global phase, represented by the diameter of the innermost pillars of $d = 250 ~nm$ and $d = 300 ~nm$. As can be seen on Figure \ref{fig:lens}a, these nominally equivalent metalenses  have different efficiencies of 75\% and 81\% (values at iteration 0), respectively. Shape optimization was applied with Fourier decomposition up to $m = 6$, and again with minimum gap of $90 ~nm$ and minimum feature of $80 ~nm$. Given the circular symmetry of the device, we only simulated one-quarter of the structure, and only one polarization (TE). Polarization insensitivity is enforced by computing the gradient for TM polarization from the TE gradient rotated by $90\deg$. The optimization results in Figure \ref{fig:lens}a show that the efficiency is substantially improved for both initial structures, reaching 88\% and 90\%. 

These results reinforce two aspects already observed for metagratings: first, shape optimization can improve the efficiency of library structures, regardless of the specific initial choice. This is not to say that every choice leads to the same final efficiency though (as is clear in Figure \ref{fig:lens}a). Second, the efficiency is improved beyond what could be achieved with such library. Figure \ref{fig:lens}a shows cuts of the z-component of Poynting vector in the xz plane (z being the propagation direction), starting at $z = 0$ (base of the pillars) and in the free-space region (above the $1 ~\mu m$ pillar height) with very little undesired diffraction observed. The beam at the focus plane is shown in Figure \ref{fig:lens}c. The full metalens structure for the $d=300 ~nm$ case is shown Figure \ref{fig:lens}d, and Figure \ref{fig:lens}e shows a zoomed region at the metalens center before and after shape optimization. As can be seen, all features have smooth boundaries, the minimum feature size is approximately $130 ~\mu m$, and respecting the $80 ~nm$ minimum gap. As outlined in Appendix \ref{appendix:math}, all values reported here represent the absolute efficiency, defined as the projection of the output field onto the target field divided by the power incident on the metasurface.

\section{Fabrication and characterization}
\label{section:experiment}

To validate the method, various metagratings were fabricated and characterized experimentally. The samples were fabricated using a conventional top-down approach, allowing for high-throughput and reproducibility. Choosing amorphous silicon as the nanopillars’ material allows for low losses and high contrast refractive index across the telecommunications c-band. The metasurfaces were fabricated on a $500~\mu m$ thick fused silica wafer, coated with $1~\mu m$ aSi using PECVD. An adhesive layer (HDMS) is spun on the wafer to promote adhesion of the negative e-beam resist (ma-N 2400) and the latter is then baked and coated with a charging dissipating solution (e-spacer). The nanopillars pattern is then exposed using e-beam lithography and developed in AZ 726 developer, while the e-spacer layer is removed in water. Exploiting a RIE (SPTS Rapier) technique, with simultaneous injection of etcher (SF6) and passivation (C4F8) gases in the chamber, the pattern is transferred onto the a-Si, using the resist as etching mask. The residual resist is then removed with oxygen plasma asher (Matrix Plasma Asher). After fabrication, the samples were characterized using a tuneable laser to measure the diffraction efficiency over a wide spectral range for both TE and TM polarizations. All values reported here represent the absolute diffraction efficiencies, defined as the ratio between the output power in the first order diffraction and the power incident on the metasurface, exactly how it is defined in the simulations (often in the literature, relative diffraction efficiencies are reported, and some times, even absolute efficiencies values don't properly consider Fresnel reflections at the glass-air interface). A detailed description of the experimental setup and procedure is provided in Appendix \ref{appendix:exp}.

\begin{figure}[b!]
	\centering
	\includegraphics{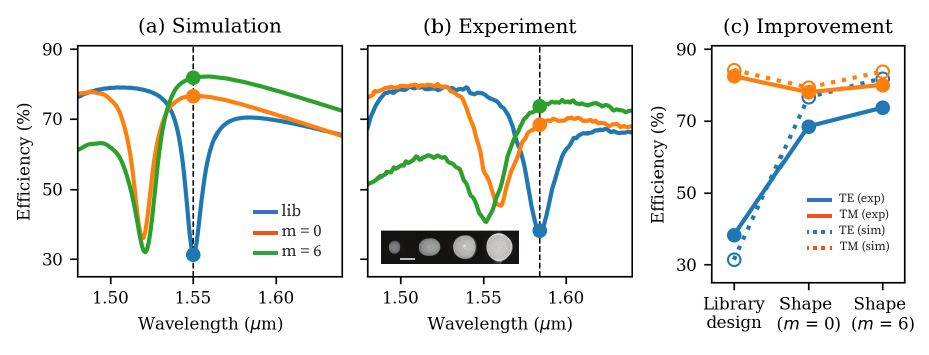}
	\caption{Simulated (a) and measured (b) TE spectral responses for the \textit{resonant} 51-degree beam deflector. Different colors show the first order diffraction efficiency for the initial and shape optimized designs with Fourier order $m = 0$ and $m = 6$ (the color legend applies to both (a) and (b) plots). A scanning electron microscope image of the fabricated structures for $m = 6$ is shown in the inset (scale bar is 200 nm). In (c), the efficiency at the design wavelength is plotted for TE and TM polarizations.}
	\label{fig:51L}
\end{figure}

Figures \ref{fig:51L}a and b show the simulated and experimental spectral responses for the \textit{resonant} $51^{\circ}$ metagrating design from Figure \ref{fig:gratings_sim}a. We only show the TE polarization as the efficiency for TM is relatively flat in this wavelength region. The initial library design (in blue) exhibits a clear resonance at $1.55 ~\mu m$. In the experiment, this resonance was slightly red-shifted to $1.58 ~\mu m$, which we attribute to small variations in the fabrication parameters such as refractive index, small sidewall angle and pillar sizes. For the sake of comparison between experiment and simulations, all efficiencies were then measured at the resonance wavelength, as indicated by the solid dots in the figures. In both, simulations and experiments, the shape optimized metagratings exhibited improved efficiencies. Comparing their spectra, we observe that shape optimization improves the diffraction efficiency by shifting by shifting the resonance away from the design wavelength. Furthermore, one can see that the shape optimization induces a tilt in the spectrum and the efficiency at the design wavelength, to the right of the resonance, is further increased. Remarkably, these spectral signatures, shift and tilt, are clearly observed in the experimental results in Figure \ref{fig:51L}b. A comparison of the efficiencies predicted in simulations and observed experimentally is shown in Figure \ref{fig:51L}c for both TE and TM, with reasonably good agreement. It is also remarkable that the experimental results show the trade-off predicted in the TE-TM efficiency for the two different Fourier orders: simple circles ($m=0$) increase the TE at the expense of TM while $m=6$ breaks this trade-off.

\begin{figure}[h!]
	\centering
	\includegraphics{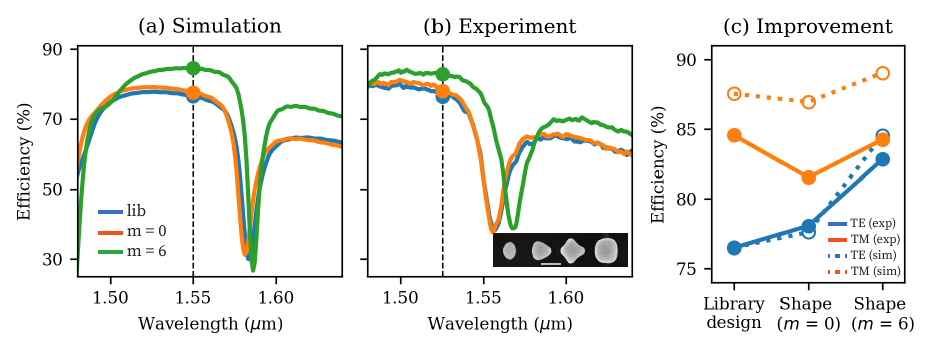}
	\caption{Simulated (a) and measured (b) TE spectral responses for the \textit{non-resonant} 51-degree beam deflector. Different colors show the first order diffraction efficiency for the initial and shape optimized designs with Fourier order $m = 0$ and $m = 6$ (the color legend applies to both (a) and (b) plots). An scanning electron microscope image of the fabricated structures for $m = 6$ is shown in the inset (scale bar is 300 nm). In (c), the efficiency at the design wavelength is plotted for TE and TM polarizations.}
	\label{fig:51H}
\end{figure}

We also fabricated the \textit{non-resonant} $51^{\circ}$ meta-grating design from Figure \ref{fig:gratings_sim}b, and the results for TE are shown in Figure \ref{fig:51H}. As mentioned, this was the highest efficiency obtained with this library and one can see that differently from the previous case, it pushed away the resonance approximately 34 nm above the design wavelength. Again the spectrum is slightly shifted in the experiment, and therefore all measurements were shifted accordingly. Differently than in the previous \textit{resonant} case, the shape optimization now does not lead to significant shifts in the resonance, and mostly push the spectrum upwards at the design wavelength. The experimental results exhibit the same behavior as predicted in simulations, and a quantitative comparison in the efficiencies is shown in Figure \ref{fig:51H}c. Similarly to the \textit{resonant} case, here we also observe that the higher Fourier order $m=6$ exhibits high efficiency for both TE and TM, reaching about 83\% and 84\% absolute efficiencies (higher than both the library design and the $m=0$ device).

\begin{figure}[b!]
	\centering
	\includegraphics{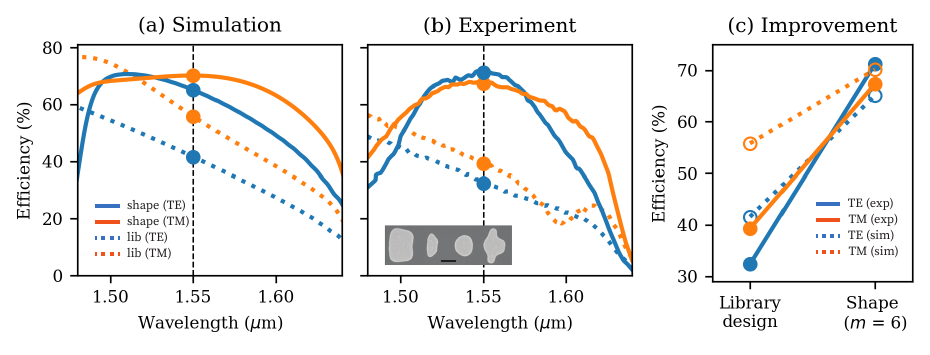}
	\caption{Simulated (a) and measured (b) spectral responses for both TE and TM polarizations for a meta-gratings at 70-degree deflection angle. Different colors show the first order diffraction efficiency for the initial and shape optimized designs with Fourier order $m = 6$ (the color legend applies to both (a) and (b) plots). An scanning electron microscope image of the fabricated structures is shown in the inset (scale bar is 200 nm). In (c), the efficiency at the design wavelength is plotted for TE and TM polarizations.}
	\label{fig:70}
\end{figure}

Finally, we then designed and fabricated a $70^{\circ}$ metagrating, showing broadband and polarization insensitive operation. The spectral efficiencies for the initial library design and for the shape optimized metagratings are shown in Figure \ref{fig:70} in dashed and solid lines, respectively. The experimental results in (b) reproduce qualitatively the simulations in (a), with measured peak efficiency above 70\%. The shape optimized meta-gratings with Fourier order $m=6$ exhibit a relatively flat spectrum centered at $1.55 ~\mu m$, with substantially higher efficiencies than the library for both TE and TM. A quantitative comparison between the simulation and experimental efficiencies is shown in Figure \ref{fig:70}c.

\section{Discussion}
\label{section:conclusions}

The shape optimization method discussed in this paper presents an alternative to conventional library approach  in metasurface design and to the more general topology optimization. As discussed, despite being simple and computationally fast, the library method falls short in many aspects, most notably by not capturing non-local coupling effects and not being amenable to general waveform incidences. Our results demonstrate that library structures exhibiting low efficiency due to the aforementioned limitations can be substantially improved with shape optimization. We show results where a resonant library structure with only 30\% TE efficiency was enhanced to 82\%, while still maintaining high TM efficiency at 84\%, smooth boundaries and respecting minimum features size and minimum gaps. Furthermore, we showed that even the highest efficiency library is also significantly improved going from 76\% to 85\% for TE, while maintaining high efficiency for TM at 90\%, and again respecting fabrication constraints. The general shape optimization explores more degrees of freedom than parametrized structures, and it is not therefore surprising that it leads to higher efficiencies. For example, parametrized aSi elliptical pillars were used to design 50 deg gratings at 900 nm wavelength and obtained absolute efficiency of 64\% for a single polarization \cite{zhou2024inverse}. Comparatively, shape optimization leads to TE and TM efficiency of 85\% and 90\%. Using rectangular pillars, a metalens with 0.78 numerical aperture operating at 850 nm was designed using parametrized rectangular aSi pillars, and obtained efficiency of 78\% (no minimum feature size or gap constraints). In contrast, shape optimized metalens with slightly higher NA of 0.83 resulted in efficiency of 90\%, including fabrication constraints.

Compared to topology optimization, shape optimization explores a reduced design space. In topology, the output structure can contain many more detailed features, and it most likely leads to higher overall efficiencies than shape optimization. For example, topology optimized gratings showed theoretical efficiencies for a 70 degrees deflection angle around 96\% (average TE/TM) \cite{sell2017large}. Comparatively, shape optimization obtained lower values of 74\% average TE/TM as shown in Figure \ref{fig:gratings_sim}c.  However, as discussed such topology optimized structures are more difficult to fabricate than shape optimized structures. Another topology optimized metagrating at 75 degrees deflection angle that was actually fabricated showed experimental efficiencies of 74\% and 75\%, for TE and TM, respectively \cite{sell2017large}. The definition of efficiency here was the power in the deflected beam normalized to the power transmitted through the bare silicon dioxide substrate. This definition differs from ours in the sense that we compute efficiency as the power in the deflected beam normalized by the power incident on the metasurface. The difference is the Fresnel reflection, which if adjusted brings the measured values to 71\% and 72\%. Comparatively, the measured efficiencies for the 70 degrees shape optimized structures from Figure \ref{fig:70} were on a similar level at 70\%. 

As a final note, the shape optimization algorithm was demonstrated here for two objective functions, i.e., TE and TM polarization, but it can be easily extended to more objectives such as multiple wavelengths or multiple angles of incidence etc. The computational cost of this optimization method is expensive, as it requires solving Maxwell equations for both forward and adjoint fields. In our optimization, we used a commercial finite-element solver in the frequency domain. Multi-wavelength optimization can greatly benefit from finite-difference time domain solvers, as the fields for all wavelengths can be obtained from only one forward and one adjoint simulations. Finally, combining this approach with GPU-accelerated FDTD can greatly speed up solving the fields \cite{hughes2021perspective, skarda2022low}. In general, the adjoint formulation tends to converge to a local optimum, with different initial conditions converging to different local optima, as was illustrated in the examples discussed here. Coupling the optimization method with parallelization is therefore beneficial to allow many starting geometries to be explored efficiently.

In summary, we presented a general shape optimization method that enables optimization of high efficiency metasurfaces. Coupled with a Fourier decomposition of the boundary gradient, the method enables high performance devices while providing greater control over the structure complexity and rigorously enforcing minimum feature size and minimum gaps. Various metagratings and metalenses were simulated, with experimental results validating the expected efficiencies. We believe these results provide a path towards manufacturability of inverse-designed high efficiency metasurfaces.

\printbibliography

\newpage

\appendix

\section{Mathematical formulation}
\label{appendix:math}

The adjoint formulation is discussed in \ref{math:adj}, including the definition of the boundary gradient. A validation of the gradient with two numerical examples is provided in \ref{math:grt_val}, and the Fourier decomposition is discussed in \ref{math:Fourier}. Finally, a generalization to an arbitrary figure of merit is discussed in \ref{math:gen_fom}. 

\subsection{Adjoint-based boundary gradient}
\label{math:adj}

\begin{figure}[b!]
	\centering
	\includegraphics{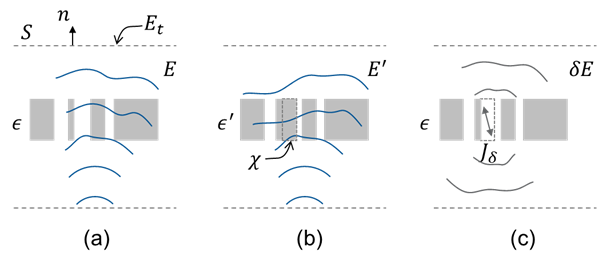}
	\caption{(a) an incident field interacts with a material configuration with permittivity $\epsilon$ to generate a field $E$ in space; (b) a new configuration $\epsilon^\prime = \epsilon + \delta \epsilon$ generates a field $E^\prime=E+\delta E$ (note that the region where the permittivity changed is denoted by $\chi$). In (c), the same field correction $\delta E$ can be produced in the initial configuration $\epsilon$ when a proper current $J_\delta$ is added in the perturbed region $\chi$.}
	\label{fig:reciprocity}
\end{figure}

The objective when designing a metasurface is to determine a specific configuration in the constituent materials, meaning their geometrical shapes and dielectric permittivity, so that an incoming field is transformed into a target field $E_t $ at the output surface $S$ (Figure \ref{fig:reciprocity}a). Suppose we start with an arbitrary configuration represented by a dielectric permittivity distribution $\epsilon$, which in general varies in space, and then change it from $\epsilon^\prime = \epsilon + \delta \epsilon$ by modifying a certain region $\chi$ within the optimization domain (Figure \ref{fig:reciprocity}b). For example, $\epsilon$ may represent an arrangement of meta-atoms (with permittivity $\epsilon_2$) embedded in a background material (with permittivity $\epsilon_1$). Enlarging one of the meta-atoms changes the permittivity in the region $\chi$ from the background value $\epsilon_1$ to the meta-atom’s permittivity  $\epsilon_2$. In general, a set of perturbations create a new configuration represented by  $\epsilon'$. The question we want to answer is where and how to change the structure so that the field changes from $E$ to  $E^\prime=E+\delta E$, moving in a direction closer to our target field $E_t$ at the output surface $S$. 

Typically, a figure of merit is used to ‘measure’ how close the field $E$ is to the target $E_t$.  For example, $E_t$ might be the mode of a waveguide positioned at the output of the domain or simply a desired field in free-space (such as a particular diffraction order in a grating or a focused Gaussian beam and so on). We can express the projection of the field $E$ into the target field $E_t$ as:

\begin{equation}
	\label{eq_app:F}
	F=\frac{1}{4\ P}\int_{S}{\left(E\times H_t^\ast+E_t^\ast\times H\right)\cdot n}da,
\end{equation}

where $n$ is the normal unit vector at the output surface S, pointing outwards from the simulation domain. If we normalize both the incoming and target fields ($P$=1\ W), then the efficiency of our device is directly obtained by $\eta=\left|F\right|^2$. Any modification in the metasurface structure that changes the fields by $\left(\delta E,\delta H\right)$ impacts the efficiency as:

\begin{equation}
	\label{eq_app:delta_eta}
	\delta\eta=2\ \mathrm{\mathrm{Re}}\left[F^\ast\delta F\right],
\end{equation}

where

\begin{equation}
	\label{eq_app:delta_F}
	\begin{aligned}
	\delta F & =\frac{1}{4\ P_t}\int_{S}{\left(\delta E\times H_t^\ast+E_t^\ast\times\delta H\right)\cdot n}da \\
		& =\frac{1}{4\ P_t}\int_{S}\left[-\left(n\times H_t^\ast\right)\cdot\delta E+\left(n\times E_t^\ast\right)\cdot\delta H\right]da.
	\end{aligned}
\end{equation}

The challenge is to determine how to modify the metasurface geometry in each iteration so $\delta\eta$ increases, as opposed to brute force trial and error every possible modification. The derivation goes as follows: we first perform two simulations to extract information in the existing medium configuration, so called forward and adjoint simulations. In the forward simulation, we calculate the field $E$ for the unperturbed medium $\epsilon$ excited by our input source. This allows us to estimate any new polarization currents $J_\delta$ that might be generated when the medium is modified by $\delta\epsilon$ (i.e., any small amount of material placed on the surface that generates a new polarization current), as illustrated in Figure \ref{fig:reciprocity}c. In the adjoint simulation, we place specific currents sources $J_a$ on the output surface $S$ that excites our medium backwards, generating a field in our domain denoted by $E_a$. We choose the currents such the adjoint field is equivalent to propagating our target field backwards into the medium. The next step is to relate these currents $J_a$ and $J_\delta$ using the reciprocity theorem. This will give us a recipe for choosing where to place $J_\delta$ (i.e. where to place $\delta\epsilon$) that ensures our efficiency increases. We now follow these steps mathematically.

The field correction $\delta E$ obtained in the perturbed medium $\epsilon^\prime$ can be generated in the ‘known’ configuration $\epsilon$ if a proper current $J_\delta$ is placed in the perturbed regions $\chi$, where:

\begin{equation}
	\label{eq_app:J_delta}
	J_\delta=j\omega\delta\epsilon\ E^\prime,
\end{equation}

This is illustrated in Figure \ref{fig:reciprocity}c. Physically, this current is the time derivative of the additional polarization in the medium $J_\delta=\partial_t\delta P=\partial_t\delta\epsilon E^\prime=j\omega\delta\epsilon E^\prime$, where we assume all fields are harmonics with dependency $e^{j\omega t}$. In other words, $J_\delta$ are the polarization currents in the medium. This can be seen by directly writing down Ampère’s law in a source free region in both perturbed and unperturbed medium:

\begin{equation}
	\label{eq_app:curl_eq}
	\nabla\times\ H=j\omega\epsilon\ E, \text{ and }
	\nabla\times\ H^\prime=j\omega\epsilon^\prime E^\prime,
\end{equation}

where again $\epsilon^\prime=\epsilon+\delta\epsilon$ and $E^\prime=E+\delta E$. Subtracting one from the other we obtain directly:

\begin{equation}
	\label{eq_app:new_curl}
	\nabla\times\delta\ H=j\omega\epsilon\delta\ E+J_\delta.
\end{equation}

\begin{figure}
	\centering
	\includegraphics{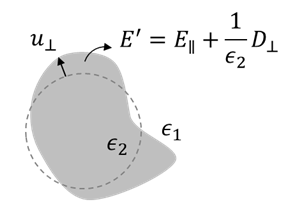}
	\caption{an initial boundary (dashed line) is deformed by a normal displacement field $u_\perp$. In regions where $u_\perp$ is positive the permittivity changes by $\delta \epsilon=\epsilon_2-\epsilon_1$, while where $u_\perp$ is negative $\delta \epsilon=\epsilon_1-\epsilon_2$. In the perturbed regions (i.e. near the boundaries), the corrected electric field can be estimated by considering the continuous tangential electric $E_\parallel$ and normal displacement $D_\perp$ fields, obtained in the original configuration.}
	\label{fig:shape_change}
\end{figure}

The last equation states that the fields $\left(\delta E,\delta H\right)$ are generated in the unperturbed medium $\epsilon$ by the current source $J_\delta$. At first this expression does not seem to be very useful because we do not yet know the perturbed field $E^\prime$ (and therefore we can’t determine $J_\delta=j\omega\delta\epsilon E^\prime$). However, note that we do not need to know $E^\prime$ everywhere in space, we only need to find an approximation for $E^\prime$ in the perturbation regions $\chi$, i.e., where $\delta\epsilon$ is non-zero. This is illustrated in Figure \ref{fig:shape_change}. The dashed line represents the original boundary of a given feature (a circular pillar in this case) that is then distorted by an amount $u_\bot$, which of course can vary along the perimeter of the shape. The tangential component of the electric field is continuous across the boundary and therefore can be considered approximately unchanged if the boundary displacement is infinitesimal (i.e. $E_\parallel^\prime=E_\parallel$ for $u_\bot\rightarrow0$). This is however not true for the normal electric field. Since it is discontinuous, the field change is finite no matter how small the boundary displacement is \cite{johnson2002perturbation}. We can circumvent this problem by noting that the normal displacement field $D_\bot$ is continuous, and therefore:

\begin{equation}
	\label{eq_app:E_prime}
	E^\prime\cong\ E_\parallel+\frac{1}{\epsilon^\prime}D_\bot,
\end{equation}

\noindent and so

\begin{equation}
	\label{eq_app:J_delta_app}
	J_\delta\cong\ j\omega\delta\epsilon\left(E_\parallel+\frac{1}{\epsilon\prime}D_\bot\right),
\end{equation}

Note that this approximation treats correctly abrupt material boundaries, i.e., with discontinuous permittivity across the boundary \cite{johnson2002perturbation}. In most topological optimization algorithms, the medium is represented by an approximate continuous permittivity, in which case the field is always continuous and therefore $E^\prime\cong E$. This means that in topology optimization, the material slowly evolves from a continuous distribution, eventually converging to a binary index distribution (i.e. manufacturable medium). This evolution however makes it difficult to control appearances of new boundaries, which can be dealt directly with shape optimization as discontinuous boundaries are naturally treated.

For the adjoint simulation, we illuminate the structure with an incident field $\left(-E_t^\ast,H_t^\ast\right)$, where the minus reflects the reversed propagation direction (i.e. propagating into the domain). This can be done through the Principle of Equivalence \cite{chen1989mathematical}, which states that any incident field into a domain can be produced by equivalent electric and magnetic  currents $J_a$ and $K_a$ on the surface $S$ enclosing this domain, where:

\begin{equation}
	\label{eq_app:adj_currents}
	J_a=-n\times H_t^\ast\text{, and } K_a=-n\times E_t^\ast,
\end{equation}

\noindent where again these surface currents produce a field in our unperturbed domain denoted by $E_a$. The interpretation of this is clear, it is just the response of the metasurface when excited by our target field instead of our original incident field. 

These considerations result in two current-field pairs with a meaningful physical interpretation: the first pair comes from the forward simulation, where we determine that polarization currents $J_\delta$ located in the ‘to-be-perturbed’ regions $\chi$ produces the field correction $\left(\delta E,\delta H\right)$. The second pair comes from the adjoint simulation, where the currents $\left(J_a,K_a\right)$ located on the domain surface $S$ produce the adjoint field $\left(E_a,H_a\right)$. The goal is now to find a relation between these two pairs that tells us where the currents $J_\delta$ should be placed so that $E+\delta E\rightarrow E_t$. This is accomplished by evoking the Lorentz reciprocity theorem \cite{landau2013electrodynamics}, which states that in a medium characterized by symmetric $(\epsilon,\mu)$, the fields $(E_1,H_1)$ and $(E_2,H_2)$, respectively produced by two different sets of localized currents $(J_1,K_1)$ and $(J_2,K_2)$, satisfy the relationship:

\begin{equation}
	\label{eq_app:recip}
	\int_{\Omega}\left(J_1 \cdot E_2 - K_1 \cdot H_2\right) dv = \int_{\Omega}\left(J_2 \cdot E_1-K_2 \cdot H_1\right) dv.
\end{equation}

Using the Dirac’s delta to represent our surface current as volume currents $\left(J_a,K_a\right)\ast\delta S$, where $\delta S$ is surface differential, the volume integrals involving the adjoint currents become surface integrals, and therefore the Lorentz reciprocity applied to our pairs of current-field becomes:

\begin{equation}
	\label{eq_app:recip2}
	\int_{S}\left(J_a \cdot \delta E-K_a \cdot \delta H\right)da = \int_{\chi} (J_\delta \cdot E_a) dv.
\end{equation}

Note that the volume integrals involving $J_\delta$ is performed only in the region $\chi$, because only there $J_\delta$ is non-zero. Furthermore, because of our choice for the adjoint currents, we can express the differential in the figure of merit as: 

\begin{equation}
	\label{eq_app:deltaF_J}
	\begin{aligned}
		\delta F & =\frac{1}{4\ P}\int_{S}\left[-\left(n\times H_t^\ast\right)\cdot\delta E+\left(n\times E_t^\ast\right)\cdot\delta H\right]da \\
				& =\frac{1}{4\ P}\int_{S}\left[J_a\cdot\delta E-K_a\cdot\delta H\right]da \\
				& =\frac{1}{4\ P}\int_{\chi}{(J_\delta\cdot E_a)}dv.
	\end{aligned}
\end{equation}

Using the expression for $J_\delta$ and writing $E_a=E_{a,\parallel}+\frac{1}{\epsilon_2}D_{a,\bot}$(which is exact), we obtain:

\begin{equation}
	\label{eq_app:deltaF_fields}
	\delta F=\frac{j\omega\delta\epsilon}{4\ P}\int_{\chi}{\left(E_\parallel\cdot E_{a,\parallel}+\frac{1}{\epsilon\ \epsilon\prime}D_\bot\cdot D_{a,\bot}\right)}dv.
\end{equation}

The reciprocity theorem allows us to predict the change in the figure of merit by performing an integration involving quantities directly at the metasurface region, and not on the output surface S. This also means that maximizing this integral in $\chi$ directly maximizes our figure of merit. The integration in the volume $\chi$ can be mapped to an integration on the surface of the original shape. Say $s$ denotes the arc length along the closed boundary $\Gamma$ encircling the nanopillar and $u_\bot$ is the normal component of the boundary displacement (Figure \ref{fig:shape_change}). The volume element is therefore $dv=u_\bot dsdz$. Assuming the shape only changes in the x,y plane (i.e. all pillars are uniform along the propagation direction z), then:

\begin{equation}
	\label{eq_app:deltaF_fields2}
	\delta F=\frac{j\omega\delta\epsilon}{4\ P}\int{u_\bot\left(E_\parallel\cdot E_{a,\parallel}+\frac{1}{\epsilon\ \epsilon\prime}D_\bot\cdot D_{a,\bot}\right)}dsdz,
\end{equation}

and the change in efficiency can be written finally as:

\begin{equation}
	\label{eq_app:delta_eta_final}
	\delta\eta=\frac{\omega\delta\epsilon}{2\ P}\ \int_{\Gamma}{(u_\bot g)}ds,
\end{equation}

where we define a gradient function $g=Re\left[jF^\ast\int{\left(E_\parallel\cdot E_{a,\parallel}+\frac{1}{\epsilon\ \epsilon\prime}D_\bot\cdot D_{a,\bot}\right)}dz\right]$, which includes an integration along the pillar height. The efficiency change is always positive if we choose $u_\bot=h\ sign\left(\delta\epsilon\right)\ g$, where $h$ is a scaling factor.

With that, the algorithm is summarized as follows:

\begin{itemize}
	\item{Initialize the structure with a set of known geometrical features (meta-atoms), with any shape. For example, a set of uniform circular pillars or a metasurface previously design with a library approach;}
	\item{Compute the forward $E$ and adjoint $E_a$ fields;}
	\item{With these fields, calculate the gradient function $g$ along the boundary of every pillar;}
	\item{Update the boundary shape by displacing every point in its perimeter by $u_\bot=h\ sign\left(\delta\epsilon\right)\ g$. Go back to step 2 until the efficiency reaches a local maximum;}
\end{itemize}

The scaling factor $h$ does not alter the deformation function $u_\bot$ but simply scales how much the overall shape is deformed. Obviously if $h$ is too large, the approximation for $E^\prime$ and thus for $J_\delta$ eventually becomes a poor one. There are many ways in which $h$ could be chosen. A simple method is to first calculate $h$ so that the efficiency change is bounded (say no more than 1\% or no less than 0.01\% or any other value). Then, one might impose the maximum of $u_\bot$ (say less than 5 nm) and re-scale $h$. A more general method to deform the boundary is using level-set functions.

\subsection{Gradient validation}
\label{math:grt_val}

\begin{figure}[h!]
	\centering
	\includegraphics{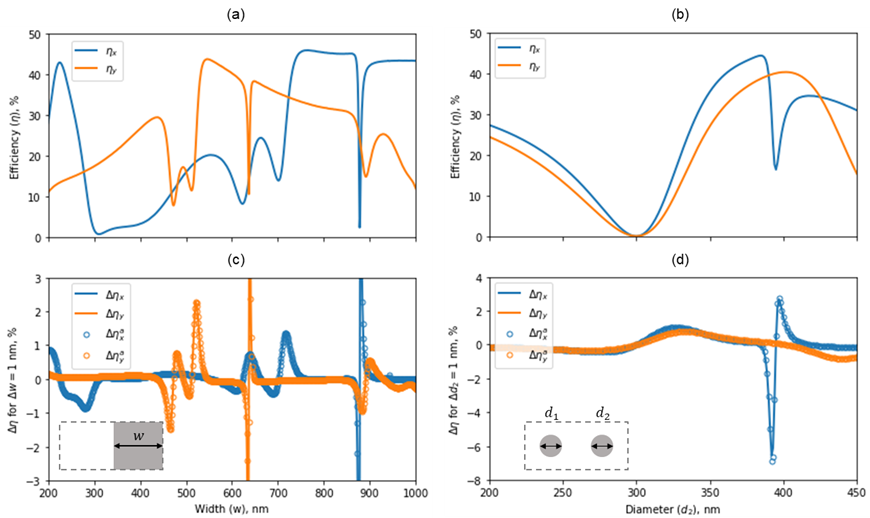}
	\caption{Validation of the boundary gradient calculation. In (a), the efficiency of a metagrating containing only one rectangular nanopillar is numerically calculated as a function of the width w, and in (b) the efficiency of a metasurface with two circular pillars is calculated as a function of the diameter of one of them ($d_2$). Here, blue and orange represent efficiency for x-polarized (along larger domain direction) and y-polarized light (along smaller domain direction), respectively. From the efficiency, the exact gradient (solid line) is calculated for the two metasurfaces and compared with the gradient obtained from the boundary shift expression (open dots). Top and bottom figures share the same horizontal axis.}
	\label{fig:gradient}
\end{figure}

To validate the gradients obtained using the adjoint formulation, we computed how the first order diffraction efficiency change as the parameters of the geometries are varied. In Figure \ref{fig:gradient}a, the diffraction efficiency is plotted as a function of the width w of a rectangular ridge for both TE (y-pol) and TM (x-pol) polarized light. The derivatives were then numerically calculated and plotted in solid lines in Figure \ref{fig:gradient}c. For comparison, the derivatives obtained directly from the adjoint gradient are shown in circles, in excellent agreement with the exact ones. In Figure \ref{fig:gradient}b, the efficiency of a metasurface that contains two circular pillars is calculated as a function of $d_2$, diameter of the second pillar (while $d_1$ is fixed). Again, the derivates computed from the efficiency curves (solid lines) match the one obtained from the adjoint gradient, as shown in Figure \ref{fig:gradient}d.

\subsection{Fourier decomposition}
\label{math:Fourier}

It is often desirable to restrict the shape to simple smooth geometries, for example to simplify fabrication. The deformation function $u_\perp$ calculated as outlined here may be very complicated, with sharp peaks or valleys. Since we are dealing with a closed boundary, any function can be expanded in terms of an appropriate basis. We chose Fourier as it easily allows restriction to smooth round structures (the arguments presented here can be generalized to any basis). The gradient function $g$ can be expanded in terms of its Fourier components:

\begin{equation}
	g=\frac{a_0}{2}+\sum_{m=1}^{\infty}{a_m\cos{m\theta}+b_m\sin{m\theta}},
\end{equation}

where

\begin{equation}
	a_m=\frac{1}{\pi}\int_{-\pi}^{\pi}{g\cos{m\theta}d\theta}, \\
	b_m=\frac{1}{\pi}\int_{-\pi}^{\pi}{g\sin{m\theta}d\theta}, \text{ and} \\
	\theta=2\pi\frac{s}{s_t}.
\end{equation}

Here, $s_t$ is the total boundary length of a given meta-atom. A circular shape is maintained if we retain only the zero-order term $a_0$: $u_\bot=h\ sign\left(\delta\epsilon\right)\ \frac{a_0}{2}$. Retaining first order terms represent slight displacements, second order terms represent elliptical distortions and so on. In general:

\begin{equation}
	u_\bot=h\ sign\left(\delta\epsilon\right)\ \left(\frac{a_0}{2}+\sum_{m=1}^{\infty}{a_m\cos{m\theta}+b_m\sin{m\theta}}\right)
\end{equation}

\noindent and therefore, the change in efficiency is:

\begin{equation}
	\delta\eta=\frac{\omega\delta\epsilon}{4\ P}\ hs_tsign\left(\delta\epsilon\right)\left[\frac{a_0^2}{2}+\sum_{m=1}^{\infty}\left(a_m^2+b_m^2\right)\right]
\end{equation}

Note that due to the orthogonality of the Fourier basis, each term contributes independently to an increase in efficiency. Therefore, we can choose to retain a finite number of terms. Finally, if we want to optimize for multiple parameters, for example TE and TM polarizations, one way is to define multiple figures of merit, each leading to its own gradient functions:

\begin{equation}
	\delta\eta_{TE} = \frac{\omega\delta\epsilon}{2\ P} \int_{\Gamma}{u_\bot g_{TE}}ds,\text{ and } \\ \delta\eta_{TM} = \frac{\omega\delta\epsilon}{2\ P} \int_{\Gamma}{u_\bot g_{TM}}ds.
\end{equation}

They can be combined in different ways, for example by using the inverse of efficiency as a weighing function $(w_{TE,TM}=1/\eta_{TE,TM})$ so that the optimization tends to balance well the performance of both polarizations, $g=\left(w_{TM}g_{TM}+w_{TE}g_{TE}\right)/\left(w_{TM}+w_{TE}\right)$. One may also define a single figure of merit as the sum of multiple figures of merit, for example to take into account different wavelengths.

\subsection{General figure of merit}
\label{math:gen_fom}

The formulation here was based on a specific definition of the figure of merit, expressed as the projection of the output field into the target field, i.e., $\eta=\left|F\right|^2$ where $F=\frac{1}{4\ P_t}\int_{S}{\left(E\times H_t^\ast+E_t^\ast\times H\right)\cdot n}da.$ This can be generalized as follows:

\begin{equation}
	F=\int_{V} f\left(E,H\right)dv,
\end{equation}

\noindent where for convenience, we normalize f so that it has units of inverse of volume. The change in efficiency is then expressed as:

\begin{equation}
	\delta F=\int_{V}\left(\frac{\partial f}{\partial E}\cdot\delta E+\frac{\partial f}{\partial H}\cdot\delta H\right)dv,\mathrm{\ and} \\
	\delta F=\frac{1}{4\ P_t}\int_{S}\left(J_a\cdot\delta E-K_a\cdot\delta H\right)da,
\end{equation}

\noindent where care must be taken to deal with derivatives with respect to complex vectorial fields. The formulation then follows identically if we choose the adjoint currents as:

\begin{equation}
	\left(J_a,K_a\right)\delta S=\left(\frac{\partial f}{\partial E},-\frac{\partial f}{\partial H}\right).
\end{equation}

\section{Experimental setup}
\label{appendix:exp}

\begin{figure} [h!]
	\centering
	\includegraphics{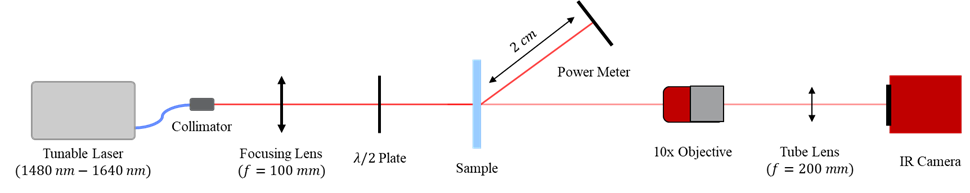}
	\caption{Schematic of the experimental setup used to measure diffraction efficiency.}
	\label{fig:setup}
\end{figure}

A schematic of the experimental setup is shown above in Figure \ref{fig:setup}. The linearly polarized output from a Santec TSL-570 tunable laser is collimated and sent through a focusing lens to create a smaller sized beam at the focal plane. A $\lambda/2$ plate is placed after the lens to control the orientation of the linear polarization. The sample is located at the focal plane of the lens, which is set to coincide with the image plane of the camera. A 10x objective and a tube lens are used to image the sample onto the camera which is used for alignment. A power meter is placed at different positions along the beam path to input power and diffracted power.

Alignment: first, the beam is aligned onto the camera without the sample in the system. Then, the position of the focusing lens is adjusted until the focus of the beam coincides with the image plane of the camera. With a focal length of 100\ mm, the focusing lens produces a beam with a diameter of approximately $100~\mu m$ at the focus for a wavelength of $\lambda$=1550\ nm. This beam size is chosen to ensure that the beam is small enough to fit entirely within one of the $250~\mu m$ x $250~\mu m$ grating patterns. Next, the polarization is set by rotating the $\lambda/2$ plate to minimize the power passing through a linear polarizer that is set to x-pol (horizontal) or y-pol (vertical), relative to the table. Then, the sample is placed at the focus of the beam in the image plane of the camera. With the sample positioned so that the beam passes through the bare glass substrate, interference fringes resulting from a tilt of the sample are visible on the camera. The sample is aligned for normal incidence by adjusting the tilt angle until these interference fringes disappear. Finally, the sample is positioned so that the beam passes through one of the grating patterns and is rotated until the height of the diffracted beam matches the height of the input beam.

Measurement Procedure: the input power is measured by placing the power meter in the beam path right before the sample. The power is measured as the tuneable laser is swept from 1480 nm – 1640 nm in steps of 0.1 nm. Then, the beam is centered onto one of the grating patterns with the help of the imaging system. The diffracted power is measured by placing the power meter in the path of the 1st-order diffracted beam, positioned as close as possible to the sample while ensuring that the 0-order beam is not detected by the sensor (2 cm). Again, the power is measured while the tuneable laser is swept over the same wavelength range.

\begin{figure} [h!]
	\centering
	\includegraphics{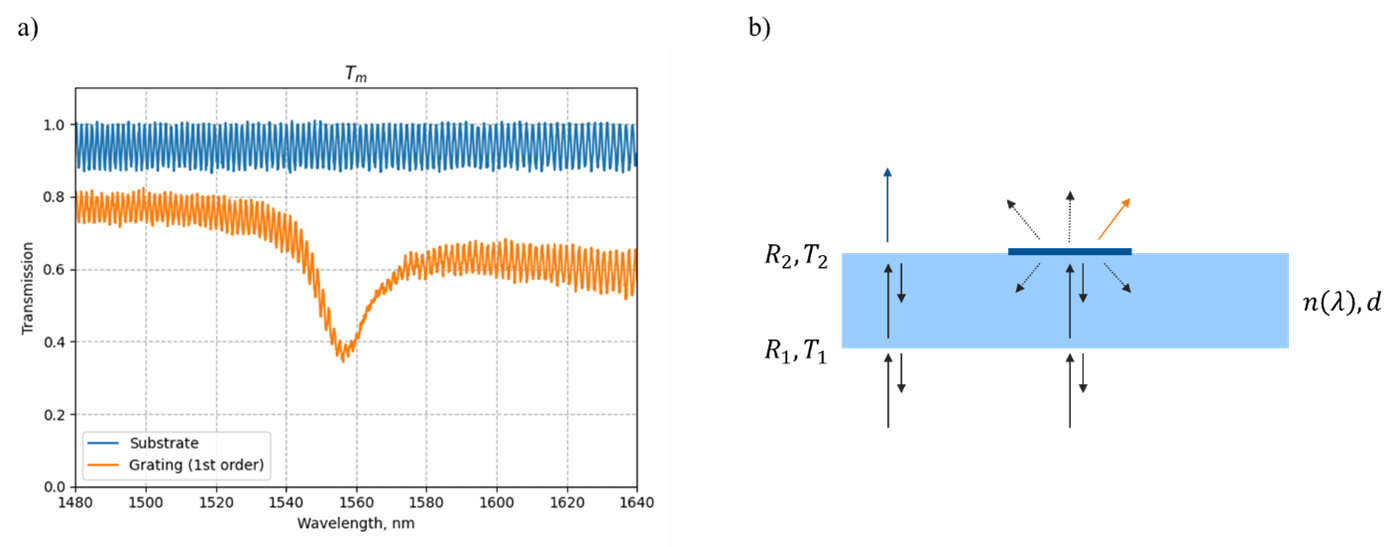}
	\caption{(a) The measured transmission through the bare substrate (blue line) and the 1st-order diffraction of one of the gratings (orange line). (b) Diagram of the Fabry-Perot model for both the bare substrate and one of the grating patterns. The blue- and orange-colored arrows represent where the transmission was measured for the bare substrate and grating, respectively.}
	\label{fig:FP}
\end{figure}

Data Processing: in our simulations, the diffraction efficiency of the device is defined relative to the power inside the substrate before interaction with the metasurface. During measurement, however, we can only measure the power before the substrate and use this as our reference. This results in many oscillations in the measured transmission due to interference within the fused silica substrate, as shown in Figure \ref{fig:FP}a. This can be understood by recognizing that the substrate acts as a Fabry-Perot cavity between the surfaces of the substrate and the 0th-order reflection from the metasurface. A diagram is shown in Figure Figure \ref{fig:FP}b. The transmission of a Fabry-Perot cavity for normal incidence is given by:

\begin{equation}
	T=\frac{T_1T_2}{1+R_1R_2-2\sqrt{R_1R_2} cos[2 k n(\lambda) d]},
\end{equation}

\noindent where $d=500~\mu m$ is the thickness of the substrate, $n(\lambda)$ is the wavelength-dependent refractive index of fused silica, $k=2\pi/\lambda$ is the wavenumber, and $\lambda=1550 ~nm$. In this equation, the parameter $T_2$ represents the transmission of purely the second interface of the sample, without any interference effects from the first interface. It is the transmission of the sample relative to the power inside the substrate, exactly how we define the diffraction efficiency. With knowledge of the transmission and reflection coefficient for the first interface ($T_1$ and $R_1$), the value of $T_2$ can be extracted from $T$, which we measure directly. Consider the maximum and minimum values of the Fabry-Perot transmission function. We find:

\begin{equation}
	T_{max}=\frac{T_1T_2}{\left(1-\sqrt{R_1R_2}\right)^2}\text{ and } \\
	T_{min}=\frac{T_1T_2}{\left(1+\sqrt{R_1R_2}\right)^2}.
\end{equation}

The values of $T_{max}$ and $T_{min}$ can be extracted from the measurements. $R_1$ and $T_1$ are the reflection and transmission coefficient for the back surface of the substrate, with values given by the Fresnel equations. For normal incidence from air to fused silica we have,

\begin{equation}
	R_1=\left|\frac{1-n\left(\lambda\right)}{1+n\left(\lambda\right)}\right|^2\text{ and }
	T_1=1-R_1,
\end{equation}
 
\noindent where $n(\lambda)$ is given by the Sellmeier equation for fused silica. By rearranging the equations for $T_{max}$ and $T_{min}$ we can find $T_2$ by first solving for the unknown quantity $\sqrt{R_1R_2}$ using the equations below:

\begin{equation}
	\sqrt{R_1R_2}=\frac{\left(1-\sqrt{\frac{T_{min}}{T_{max}}}\right)}{\left(1+\sqrt{\frac{T_{min}}{T_{max}}}\right)} \text{ and } \\
	T_2=\frac{T_{max}\left(1-\sqrt{R_1R_2}\right)^2}{T_1}.
\end{equation}

To verify this procedure for extracting $T_2$, we first consider the case of a beam passing through the bare substrate for which $T_2$ can be readily calculated according to the Fresnel equations. The results are shown below in Figure \ref{fig:glass_only}.

Using the procedure outlined above for the case of transmission through the bare substrate, we find an average value of $T_{2,measured} = 0.9670079$ which agrees very well with the value given by the Fresnel equations $T_{2,Fresnel}=0.9670084$. The same processing is used for the 1st-order transmission measurements of the meta-gratings to calculate the diffraction efficiency. An example is shown below in Figure \ref{fig:grating_only}. 

\begin{figure}
	\centering
	\includegraphics{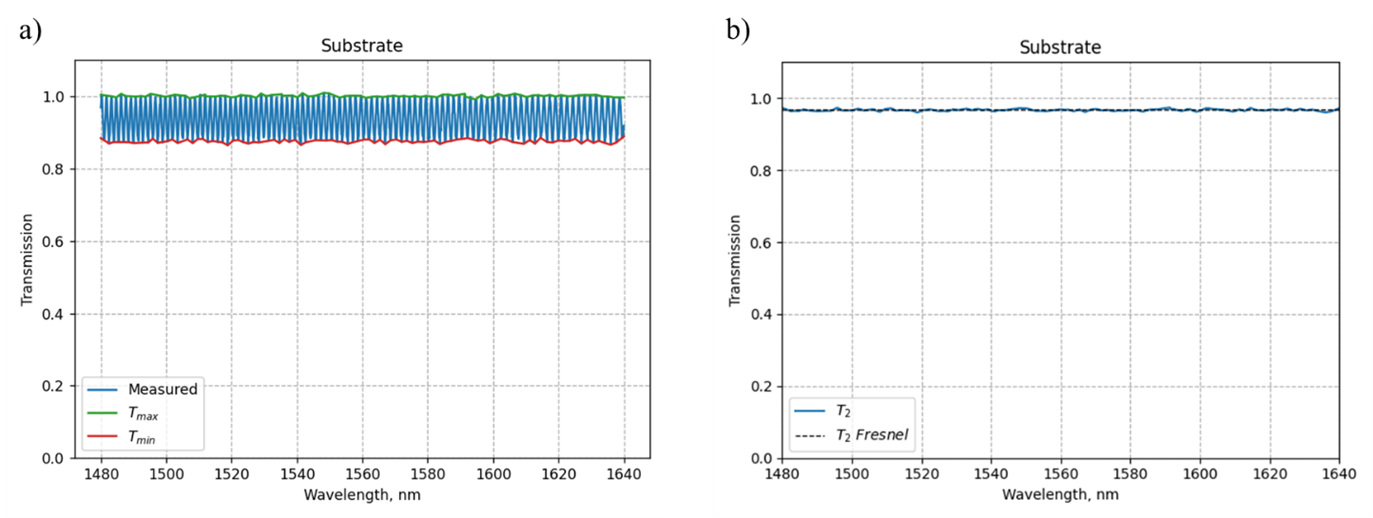}
	\caption{(a) The values of $T_{max}$ (green curve) and $T_{min}$ (red curve) are interpolated from the measured transmission (blue curve) through bare substrate. (b) The extracted value of $T_2$(blue curve) is compared with the exact value given by the Fresnel equations (dashed black line).}
	\label{fig:glass_only}
\end{figure}

\begin{figure}
	\centering
	\includegraphics{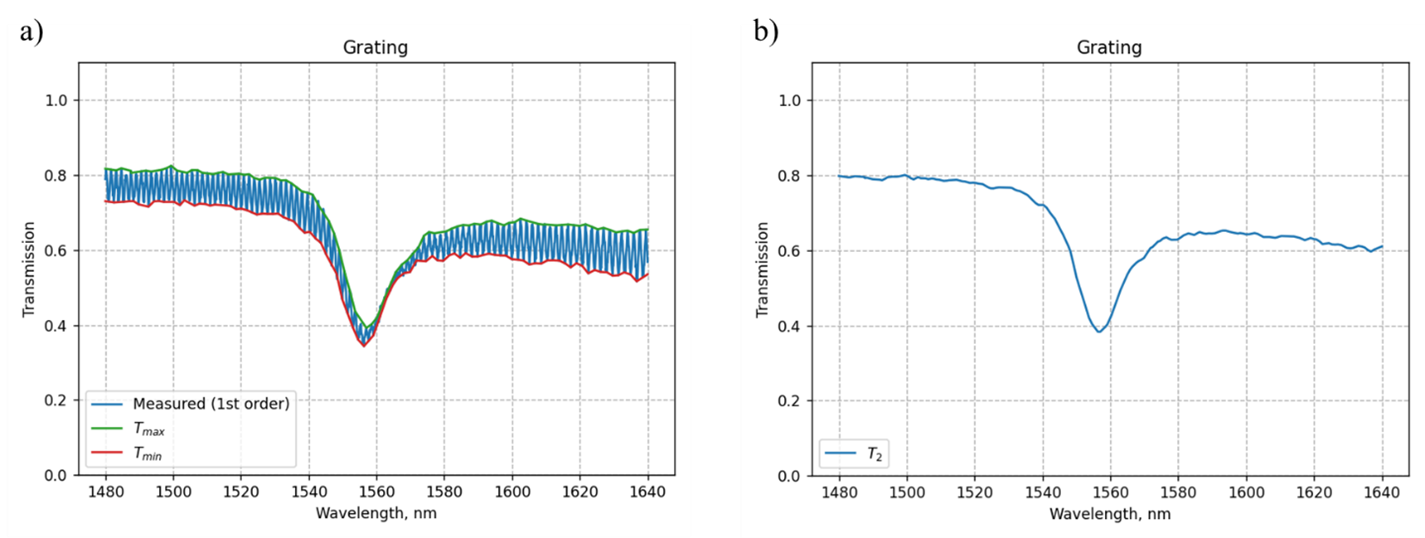}
	\caption{: (a) The values of $T_{max}$ (green curve) and $T_{min}$ (red curve) are interpolated from the measured 1st-order transmission (blue curve) of a meta-grating. (b) The extracted value of $T_2$ (blue curve), defining the measured diffraction efficiency.}
	\label{fig:grating_only}
\end{figure}

\end{document}